\documentstyle[art12]{article}
\newcommand{\be}{\begin{equation}}
\newcommand{\ee}{\end{equation}}
\newcommand{\ba}{\begin{eqnarray}}
\newcommand{\ea}{\end{eqnarray}}
\newcommand{\baa}{\begin{eqnarray*}}
\newcommand{\eaa}{\end{eqnarray*}}
\newcommand{\bb}{\begin{thebibliography}}
\newcommand{\eb}{\end{thebibliography}}
\newcommand{\ci}[1]{\cite{#1}}
\newcommand{\bi}[1]{\bibitem{#1}}
\newcommand{\lab}[1]{\label{#1}}
\newcommand{\re}[1]{(\ref{#1})}
\newcommand{\Tr}{\mbox{Tr\,}}
\newcommand{\bit}{\begin{itemize}}
\newcommand{\eit}{\end{itemize}}

\newcommand\s{{\scriptscriptstyle S}}

\begin{document}
\thispagestyle{empty}

\hfill {\sc UPRF--91--316} \qquad { } \par
\hfill {\sc CPT--91/P.2629} \qquad { } \par

\vspace*{15mm}
\begin{center}
\renewcommand{\thefootnote}{\fnsymbol{footnote}}
{\LARGE INFRARED FACTORIZATION, \\[10mm] WILSON LINES \\[10mm]
AND THE HEAVY QUARK LIMIT}

\vspace{12mm}

{\large G.~P.~Korchemsky}%
\newcommand{\st}{\fnsymbol{footnote}}%
\footnote{INFN Fellow}
\footnote{On leave from the Laboratory of Theoretical Physics,
          JINR -- Dubna}

\medskip

{\em Dipartimento di Fisica, Universit\`a di Parma and \par
INFN, Gruppo Collegato di Parma, I--43100 Parma, Italy} \par
{\tt e-mail: korchemsky@vaxpr.cineca.it}

\bigskip

{and}

\bigskip

{\large A.~V.~Radyushkin}%
${}^{\st}$

\medskip

{\em Centre de Physique Theorique, CNRS - Luminy, \par
F 13288 Marseille, France}                        \par
{\tt e-mail: radyush@cebaf}

\end{center}

\vspace*{20mm}

\begin{abstract}
It is shown that,
in QCD,  %
the same universal function
$\Gamma_{cusp}(\vartheta, \alpha_\s)$
determines the infrared behaviour of the on-shell quark form factor,
the velocity-dependent anomalous dimension in  the
heavy quark effective field theory (HQET) and
the renormalization properties
of the vacuum averaged Wilson lines with a cusp.
It is demonstrated that a combined use
of the methods
developed in the relevant different branches of
quantum field theory essentially
facilitates the all-order study of the asymptotic and analytic
properties of this function.

\end{abstract}

\newpage

\setcounter{page}1

\renewcommand{\thefootnote}{\arabic{footnote}}
\setcounter{footnote}{0}

\subsection{Introduction}
A considerable effort is being made now to construct
a new approach to the heavy-quark physics.
The basic idea \ci{Vol,Isg} is to use the fact that
the masses of the heavy quarks
$c$ and $b$ are essentially larger than the QCD scale $\Lambda$,
and to start (when appropriate) with an effective
field theory \ci{Vol,Isg}, in which the heavy quark masses
are infinite: $m_Q \to \infty$.

In the infinite mass limit,
the components of the heavy quark momentum $p_{\mu}$
tend to infinity together with the quark mass: $p \sim m_Q$,
the ratio $p_{\mu}/m_Q$ approaching a fixed value $v_{\mu}$.
The 4-velocity
$v_{\mu}  = (1/\sqrt{1-{\bf v}^2}, {\bf v}/\sqrt{1-{\bf v}^2})$
just specifies the direction in which the
infinitely heavy quark is moving,
and
the heavy quark effective theory (HQET) \ci{Eich,Georgi}
has a peculiar property
that each effective quark field $h_v(x)$ is
characterized by a 4-velocity vector $v_{\mu}$.

The electroweak interactions can change the
velocity of the heavy quark $v_1 \to v_2$ and,
as noted by Isgur and Wise \ci{Isg},
it is the 4-velocity change rather
than the momentum transfer that is an important
dynamic variable for the heavy meson form factors,
both for elastic (like $\gamma^* D \to D$ and $\gamma^* B \to B$)
and for weak ($B \to De\nu$-type)  transition form factors.
In other words, the QCD currents $\bar Q_2 \Gamma Q_1$
are substituted
in the HQET by the effective
currents $\bar h_{v_2} \Gamma h_{v_1}$
containing fields
related to the two velocities $v_1$ and $v_2$.

An important observation made in \ci{Uni}
is that the effective currents $\bar h_{v_2} \Gamma h_{v_1}$
have a universal (the same for all $\Gamma$-matrices) velocity-dependent
anomalous dimension $\gamma_w(\alpha_\s)$.
The velocity change is characterized
by the scalar product
$w \equiv (v_1 v_2)$ or,
since the velocity vectors have a unit ``length'', $v_1^2=v_2^2=1$,
by the angle $\vartheta$ between them:
$w  = \cosh\,\vartheta$.

As a matter of fact, the authors of \ci{Uni} are not the
first who encountered an angle-dependent anomalous dimension
in their calculation. Similar things happened before in
two other and,  at first sight,
entirely different branches of quantum field theory.

First, more than 10 years ago, Polyakov \ci{Pol},
trying to reformulate the dynamics of gauge fields
in terms of string operators, has observed
that the vacuum average of a Wilson loop
having a cusp, possesses extra ultraviolet
divergences. These specific divergences, ``generated''
by the cusp, are multiplicatively renormalized \ci{peo}, and the
relevant anomalous dimension
$\Gamma_{cusp}(\vartheta, \alpha_\s)$
depends on the cusp angle $\vartheta$.

The second example is related to the
infrared (IR) behaviour of perturbative QCD
which is known to be controlled
by the subprocesses involving soft gluons.
Since the soft gluons cannot change the momenta $p_i$
of the incoming and outgoing particles,
one faces here a situation completely
similar to that in the heavy-quark limit: the
particles are just moving in fixed directions
(specified by $p_i$) through the gluonic cloud.
The study of the IR behavior of perturbative QCD
has not a purely academic interest:
even if the IR singularities cancel,
there remain finite IR-induced contributions.
In our papers \ci{IR,IR1}, we
developed an approach that enables one to
study these
contributions
using a renormalization group equation.
The relevant anomalous dimensions
$\Gamma_{IR}(\vartheta_{ij},\alpha_\s)$ depend just
on the angles $\vartheta_{ij}$ between the relevant
momenta:
$\cosh\,\vartheta_{ij} =$ $(p_i p_j )/ \sqrt{p_i^2p_j^2}.$

The IR-induced terms
are especially important for high momentum
transfer processes, when
$Q_{ij}^2\equiv -(p_i-p_j)^2 \gg p_i^2 \sim p_j^2 \equiv p^2$,
{\em i.e.,} when some angles $\vartheta_{ij}$ are large.
We established that, in this  limit,
the anomalous dimension $\Gamma_{IR}(\vartheta_{ij},\alpha_\s)$,
to all orders in $\alpha_\s$,
is linear in $\vartheta_{ij}$ or, what is the same, in
$\log(Q_{ij}^2/p^2)$:
\be
\Gamma_{IR}(\vartheta_{ij},\alpha_\s)=
K(\alpha_\s) \log(Q_{ij}^2/p^2) +A(\alpha_\s).
\lab{gammaIR}
\ee
The function $K(\alpha_\s)$ plays an important role in
the IR-induced and the double logarithmic contributions
\footnote{A
          function equivalent to $K(\alpha_\s)$ is the basic object
          in another RG-type approach to the double-logarithmic
          effects developed by Collins \ci{Col}}
to different hard processes:
the same combination
\be
K(\alpha_\s)=\frac{\alpha_\s}{\pi}C_F +
            \left(\frac{\alpha_\s}{\pi}\right)^2
            C_F\left(C_A\left(\frac{67}{36}-\frac{\pi^2}{12}\right)
            -N_f\frac{5}{18}\right)
\label{kalpha}
\ee
(where $C_F=4/3$ and $C_A=3$ are
quark and gluon color factors
and $N_f=3$ is the number of light quark flavours)
governs the $x \to 1$ asymptotics of the deep inelastic structure
functions \ci{IR1} and the large  perturbative corrections to the
Drell-Yan cross section
\ci{D-Y} and the
$e^+e^--$annihilation \ci{x->1} .

One might ask: are all these anomalous dimensions independent
or there exists a
connection between all of them?
It is our goal in the present paper to study the
interrelation between the
infrared behaviour of perturbative QCD and the
renormalization properties
both of the Wilson loops (lines)  and of the effective
heavy quark currents in the HQET,
to demonstrate that the apparently unrelated functions,
each playing a fundamental role in its field,
are precisely identical and correspond to the
QCD analogue of the bremsstrahlung function known
from the early QED years.

\subsection{Infrared singularities and the heavy quark limit}
A simple but essential fact is
that there are no velocity-
or angle-dependent ultraviolet divergences in any standard
QCD diagram. This means that the
HQET
has extra UV divergences:
the heavy quark limit $m_Q \to \infty$ is singular.
In particular, a logarithmic divergence in the HQET may result from
a $\log(m_Q/\lambda)$-contribution in QCD, with
$\lambda$ remaining finite in the heavy quark limit.
However, since all the momenta $p_{\mu}$ scale as $m_Q$
in this limit:  $p_{\mu} \to m_Qv_{\mu}$,
the only combination that might be much smaller than
$m_Q^2$ is $(p^2-m_Q^2)$,  the heavy quark off-shellness,
and the singularity may result only from a
term like $\log(m_Q^2/(p^2-m_Q^2))$.
Such a term is a standard infrared logarithm, with
$(p^2-m_Q^2)$ serving as an infrared cut-off.
This indicates that the velocity-dependent divergences should be
closely related to the IR behaviour of perturbative QCD.
Indeed, studying
the hadrons containing a
single heavy quark,
one is interested in the kinematic situation when the
heavy quark energy differs from its mass by some finite amount
$E$, which has a meaning of the average energy of the light quark(s),
binding energy, {\em etc.}
Thus, the virtuality of the heavy quark $(p^2-m_Q^2) \approx 2m_QE$
is much smaller than its mass squared:
$$
\Delta= \frac{p^2-m_Q^2}{m_Q^2} \, \ll \, 1 .
$$
In the heavy quark limit, $m_Q \to \infty$,
the parameter $\Delta$ tends to zero  and,
in this sense, the heavy quark is almost on-shell.

It is well-known that,
in gauge theories,
the on-shell form factors
have infrared (IR) singularities.
The classic example is the
elastic electron on-shell ($p_1^2=p_2^2=m^2$) form factor in QED.
At the one-loop level one has, for the Dirac form factor
$$
F(Q^2,m^2) = 1- \frac{\alpha}{2\pi}
\log\left(\frac{m^2}{\lambda^2}\right)
   B\left(\frac{Q^2}{m^2}\right) +
(IR \ regular \ terms) \  ,
$$
where $Q^2=-(p_1-p_2)^2$ is the transferred momentum,
$\lambda$ is the (fictitious) photon mass
serving as  the IR cut-off parameter and
$B\left(\frac{Q^2}{m^2}\right)$ is the
bremsstrahlung function
\be
B\left(\frac{Q^2}{m^2}\right)=
\frac{1+\frac{2m^2}{Q^2}}{\sqrt{1+\frac{4m^2}{Q^2}}}
\log \left (\frac{\sqrt{1+\frac{4m^2}{Q^2}}+1}
{\sqrt{1+\frac{4m^2}{Q^2}}-1} \right ) -1 \ .
\lab{brems}
\ee
Since $\lambda^2$ is treated as a scale much smaller than $m^2$,
we have the same kinematic situation as in the heavy quark limit.

The IR logarithms in QED are known to exponentiate when summed
over all orders \ci{Fra}
\be
F(Q^2,m^2) =
\exp\left\{-\frac{\alpha}{2\pi}\log\left(\frac{m^2}{\lambda^2}\right )
B\left(\frac{Q^2}{m^2}\right)\right\}
\times (IR \ regular \ terms ).
\label{eq:dirac}
\ee
In this expression, one
can easily factorize the ``long distance''
contribution
$
\left (\mu^2/\lambda^2 \right )^{-\frac{\alpha}{2\pi}B}
$
absorbing all the IR divergences,
and the ``short-distance'' factor
$
\left (m^2/\mu^2 \right )^{-\frac{\alpha}{2\pi}B}.
$
As a result,  one can rewrite
eq.\re{eq:dirac} as
\be
F(Q^2,m^2) =
\left (\frac{\mu^2}{\lambda^2} \right )^{-\frac{\alpha}{2\pi}B}
\, C\left (\frac{m^2}{\mu^2},\frac{Q^2}{m^2}\right )
\label{eq:mudirac}
\ee
where the IR regular factor
$C(\frac{m^2}{\mu^2},\frac{Q^2}{m^2})$
now has an explicit dependence on the
IR factorization scale $\mu$.

The derivation of this factorized
representation has a striking resemblance to the standard
factorization procedure for the collinear
mass singularities applied to deep inelastic scattering
(in the form of the operator product expansion, OPE) and
to other hard processes.
Using it, one can factorize the original function, {\em e.g.,}
$W_n(Q^2, p^2)$,
the  $n$-th moment of the deep inelastic structure function,
into the short-distance coefficient function
$C_n(Q^2/\mu^2)$
and the long-distance sensitive matrix element
$f_n(\mu^2/p^2)$ absorbing all the collinear logarithms
$\log\,p^2$ singular in the $p^2 \to 0$ limit:
$$W_n(Q^2, p^2) = f_n(\mu^2/p^2)\, C_n(Q^2/\mu^2).$$

The factorization
scale $\mu$ separates the high- and low-momentum regions:
the essential momenta $k$ are $k>\mu$ for the
$C$-function and $k<\mu$ for the $f$-function.
An important
fact
is that the matrix elements of the
local composite twist-2 operators $O_n$, producing $f_n$, possess
extra UV divergences absent in the original theory (QCD).
Of course, the factorization procedure automatically imposes
the  UV cut-off $k<\mu$, and there are no actual infinities.
But now one can make use of the fact that the $\mu$-dependence
of the $f_n$-function is just the dependence on the UV cut-off
parameter and incorporate the renormalization group to study it.
That is how the $n$-dependent anomalous dimensions
$\gamma_n(\alpha_\s)$
of the composite operators $O_n$ and/or $z$-dependent
anomalous dimensions (evolution kernels) $P(z,g)$ of the
parton distribution functions $f(x, \mu^2)$ come into play.

The analogy between
the factorization properties of the QED formula \re{eq:mudirac}
and the OPE-type factorization has a very deep reason \ci{IR}:
in QCD (and in any similar gauge theory)
one can factorize out
the IR-singular terms of the on-shell quark form factor
$$
F(Q^2,m^2)= F_{IR}\times(IR\ regular\ terms).
$$
The IR-sensitive factor $F_{IR}$ accumulates all the effects due
to interactions of the massive particle (heavy quark, if one studies
the heavy quark limit) with soft quanta -
gluons and massless quarks. The explicit expression
for $F_{IR}$ is given by the vacuum average
of the Wilson line operator $W(C)$
\be
F_{IR}=\langle 0| {\cal T}
\exp\left(ig\int_0^{+\infty} dt\,v_2^\mu A_\mu(v_2t)\right)
\exp\left(ig\int_{-\infty}^0 dt\,v_1^\mu A_\mu(v_1t)\right)
|0\rangle
\equiv \langle 0| W(C) |0\rangle
\lab{IRfactor}
\ee
calculated along the classic path $C$ of the massive (``heavy'')  particle.
It goes from $-\infty$ along the direction of the initial
momentum $p_1=mv_1$ to the interaction point $0$ and then along
the direction of the final momentum $p_2=mv_2$ to $+\infty$.
The soft gluons coupling directly to the heavy particle are
described
by the gauge potentials $A_{\mu}(x)$.

To summarize,
in this limit (one can
treat it either as the IR limit $\lambda \to 0$ or
as the heavy quark limit $m \to \infty$),
all the effects due to interactions
of the massive particle with the gluons
can be described by a Wilson line.
This result has a natural interpretation:
the wave function of a very massive
particle propagating through the gluonic cloud
acquires only a phase factor
equal to the Wilson line.
All other effects are
suppressed by powers of $\lambda/m$.
In particular, they cannot
produce contributions singular in the
IR limit $\lambda \to 0 $.

Just like in the OPE case, the factorization
scale $\mu$ works as an
ultraviolet cut-off for $F_{IR}$, since only small momenta $k<\mu$
contribute to it. And such a cut-off is really necessary because, as
established by Polyakov \ci{Pol},
any Wilson loop with a cusp has extra UV divergences.

\subsection{Wilson lines and the heavy quark effective field theory}
The properties of the Wilson lines and loops
where studied in detail at the beginning of the 80's.
A very important result \ci{Ger} is that one
can interpret the  vacuum average of a
Wilson line $W(C)$ as the propagator
of one-dimensional fermions ``living'' on the integration
path $C$:
\be
\Tr W(C)= \langle \psi_a(+\infty) \bar{\psi}_a(-\infty) \rangle
\equiv \int {\cal D}\bar{\psi}(t){\cal D}\psi(t)\
\psi_a(+\infty) \bar{\psi}_a(-\infty) \exp(-iS)
\lab{1-dim}
\ee
where the action $S,$ in our case, is defined by
\be
S=i\int_{-\infty}^{+\infty}dt\,
\bar{\psi}(t)\frac{\partial}{\partial t}\psi(t)
-g\int_{-\infty}^0dt\,\bar{\psi}(t)v_1^\mu A_\mu(v_1t)\psi(t)
-g\int_0^{+\infty}dt\,\bar{\psi}(t)v_2^\mu A_\mu(v_2t)\psi(t).
\lab{action}
\ee
It is easy to show that the one-dimensional fermions defined
in this way exactly coincide with the
effective  heavy quark field of the HQET:
\be
\langle 0|h_{v_1}(v_1t)|h,v_1\rangle
=\langle \psi(t) \bar{\psi}(-\infty)\rangle
={\cal T}\exp\left(ig\int_{-\infty}^t d\sigma\,
 v_1^\mu A_\mu(v_1\sigma)\right),
\quad t < 0
\lab{equiv1}
\ee
and
\be
\langle h,v_2|\bar{h}_{v_2}(v_2t)|0\rangle
=\langle \psi(+\infty)\bar{\psi}(t)\rangle
={\cal T}\exp\left(ig\int_t^{+\infty} d\sigma\,
 v_2^\mu A_\mu(v_2\sigma)\right),
\quad t > 0 \ .
\lab{equiv2}
\ee
The representation \re{1-dim} is very convenient to
analyze the general renormalization properties of
the Wilson lines. In particular, it was found \ci{peo,Ger}
that for a smooth integration
path all the UV divergences can be compensated by
a renormalization of fields and vertices.

The Wilson lines in the r.h.s. of eqs.\re{equiv1} and \re{equiv2}
are not closed, and their end-points
produce extra UV divergences. They can be
multiplicatively renormalized, and the relevant
anomalous dimension $\gamma_{end}(\alpha_\s)$
was calculated, in the Feynman gauge, to one loop
in \ci{peo,Ger} and to two loops in \ci{Kna}
$$
\gamma_{end}(\alpha_\s) = -\frac{\alpha_\s}{\pi}C_F
 -\left(\frac{\alpha_\s}{\pi}\right)^2 C_F
  \left(\frac{19}{12}C_A - \frac{1}{3}N_f \right) .
$$
{}From the relations \re{equiv1} and \re{equiv2} it follows
that $\gamma_{end}(\alpha_\s)$ coincides with the anomalous dimension of
the effective quark field in the HQET and with that
of the one-dimensional
fermions in the theory defined by \re{action}.
The HQET calculation was performed in \ci{Gro}.

As mentioned earlier, each cusp of the integration path
also generates extra
UV divergences governed by
the  anomalous dimension $\Gamma_{cusp}$ that depends on the
cusp angle $\vartheta$.
In the theory of one-dimensional fermions \re{action} one obtains
$\Gamma_{cusp}$ as the anomalous dimension of the composite
operator $\bar{\psi}(+0)\psi(-0).$
Using the relations \re{equiv1} and \re{equiv2} one can identify
this operator with a heavy quark current
$\bar h_{v_2}(0) h_{v_1}(0)$.
Thus, the anomalous dimension of the heavy quark
current in the HQET coincides
with the cusp anomalous dimension of the relevant Wilson line
\be
\Gamma_{cusp}(\vartheta, \alpha_\s ) = \gamma_w(\alpha_\s) ,
\lab{eqHQET}
\ee
provided, of course, that $w= \cosh \, \vartheta.$

In general, it is a difficult problem to calculate Wilson loops,
even in perturbation theory.
However, it is possible to evaluate $\langle 0|W(C)|0\rangle=F_{IR}$
using the explicit form of the contour $C$ defined by eq.\re{IRfactor}.
The path $C$ has an infinite length and, as a consequence,
$\langle 0|W(C)|0\rangle$ has infrared divergences.
Moreover, $\langle 0|W(C)|0\rangle$ depends on the directions of the
rays forming
$C$. Note now that the directions defining the line $C$
are specified by the dimensionless
ratios $v_1^{\mu}= p_1^{\mu}/m$ and $v_2^{\mu}= p_2^{\mu}/m$ .
Hence, the 
dimensionful parameters that explicitly appear in the
vacuum average \re{IRfactor} are the IR cut-off $\lambda$ and
the factorization scale $\mu$, the latter, as emphasized earlier,
working as an ultraviolet cut-off
for the new divergences generated by the cusp
of the path $C$ at the point $0$.
This means that the UV scale $\mu$ and
the IR scale $\lambda$ should
basically
appear in the ratio.
There is also the ordinary renormalization
scale $\mu_R$ fixing the definition of the
running coupling constant $g(\mu_R)$:
$$
F_{IR} =  F_{IR} \left ( \frac{\mu}{\lambda}, \
 \frac{\mu}{\mu_R}, \ (v_1v_2), \  g(\mu_R) \right ) =
F_{IR} \left ( \frac{\mu}{\lambda}, \
 \frac{\mu}{\mu_R}, \ \frac{(p_1p_2)}{m^2}, \  g(\mu_R) \right ).
$$
After taking $\mu=\mu_R,$ the IR factor $F_{IR}$, being equal
to the vacuum average of a Wilson loop, obeys, in general, the RG
equation
$$
\left (\mu\frac{\partial}{\partial\mu}
    +\beta(g)\frac{\partial}{\partial g}
+ \Gamma_{IR}\left(\frac{(p_1p_2)}{m^2}, g\right) \right )
F_{IR} \left(\frac{\mu}{\lambda},1,
\frac{(p_1p_2)}{m^2}, \, g\right) = 0 \, ,
$$
where the anomalous dimension $\Gamma_{IR}$ controls the IR induced
logarithms $\log(\mu/\lambda).$ Since the path $C$ has a cusp, one
identifies $\Gamma_{IR}$ with the cusp anomalous dimension:
\be
\Gamma_{IR}\left(\frac{(p_1p_2)}{m^2}, \alpha_\s\right)=
\Gamma_{cusp}(\vartheta,\alpha_\s).
\label{eqircusp}
\ee

Furthermore, since the IR singularities of $F_{IR}$
coincide with those of the original amplitude ({\em
e.g.,}
with those present in the
on-shell form factor),
the cusp anomalous dimension $\Gamma_{cusp}(\vartheta,\alpha_\s)$
must reproduce the bremsstrahlung function \re{brems}:
\be
\Gamma_{cusp}(\vartheta,\alpha_\s)=\frac{\alpha_\s}{\pi}C_F
B\left(\frac{Q^2}{m^2}\right) +{\cal O}(\alpha_\s^2)
\lab{eqIR}
\ee
The one-loop result for the cusp anomalous dimension
can be taken from
the pioneering paper by Polyakov \ci{Pol}, where it was
calculated in the Euclidean space.
In Minkowski case, one should change $\vartheta \to i\vartheta$
to get
$$
\Gamma_{cusp}^{1-loop}(\vartheta,\alpha_\s)
=\frac{\alpha_\s}{\pi}C_F(\vartheta \coth \vartheta-1) .
$$
To check the
equivalence relation \re{eqIR}, one should
just
use
that $Q^2 =2m^2(\cosh \,\vartheta-1).$
Taking the 1-loop expression for $\gamma_w(\alpha_\s)$ from \ci{Uni},
\be
\gamma_{w}(\alpha_\s) = \frac{\alpha_\s}{\pi}C_F
\left(\frac{w}{\sqrt{w^2-1}}\log(w+\sqrt{w^2-1})-1\right)
\label{gammaw1loop}
\ee
one can also verify, at the one-loop level,
the relation \re{eqHQET}.

The equivalence relations \re{eqHQET} and \re{eqIR} manifest
a close connection between the HQET and the
theory of Wilson lines.
The cusp anomalous dimension appears here as a new
universal function  controlling the IR-induced
properties of perturbative QCD.
All the specifics of the process under study is
contained in the dependence of $\Gamma_{cusp}(\vartheta, \alpha_\s)$
on the cusp angle $\vartheta$ determined entirely by the
kinematics of the process.

\subsection{Non-abelian exponentiation theorem}
Using effective fields, the
one-dimensional fermions or the effective heavy-quark fields,
one can essentially simplify a general study of the
renormalization properties of the Wilson lines.
However, working within  the effective field theory approach,
one can easily miss some properties
of the relevant anomalous dimensions which are evident if  one treats
them in the context of the Wilson line formalism.

For example, in QED,
it is possible to get an exact expression
for $\Gamma_{cusp}(\vartheta,\alpha)$. The reason is that,
in an abelian gauge theory without massless fermions
(electrons in QED are treated as massive particles!),
the  vacuum average of
a Wilson exponential is an exponential of the vacuum average
corresponding to the photon propagator:
$$
\langle 0| {\cal T} \exp\left(ig\int_{C} dz^{\mu}
\, A_\mu(z)\right) |0\rangle =
\exp\left(\frac{(ig)^2}{2}\int_{C}dz_1^{\mu}\,\int_{C}dz_2^{\nu}
\langle 0| {\cal T} A_\mu(z_1) A_\nu(z_2) |0\rangle \right) .
$$
As a result, the cusp anomalous dimension
in QED is completely determined
by the first loop:
$$
\Gamma_{cusp}^{QED}(\vartheta,\alpha)=\frac{\alpha}{\pi}
    B\left(\frac{Q^2}{m^2}\right).
$$
There are no higher order corrections. From the IR side,
this property is well-known: all the IR singularities
are given in QED by the exponential of the one-loop result \ci{Fra}.

In QCD, the  situation is more complicated.
First, there are light quarks $u,d$
and $s$ which can be treated as
massless at a typical hadronic scale
$\sim 1\,GeV$.
Second, the gluons
are described by non-abelian gauge fields.
Nevertheless, it is possible to prove a
non-abelian exponentiation theorem \ci{Gat}:
the vacuum average of a Wilson line operator
in QCD is an exponential of
some expression, to which, of course, not
only the first loop contributes.
Still, the property observed in QED is partially preserved:
the higher-order corrections to $\Gamma_{cusp}(\vartheta,\alpha_\s)$ do
not repeat information contained in the one-loop result
or, in general, in
all the
previous loops: calculating
$\Gamma_{cusp}(\vartheta,\alpha_\s)$,  it is sufficient
to consider only the diagrams which do not contain
the lower-order ones as subgraphs, {\em i.e.,}
the essential diagrams are two-particle irreducible
in the current channel.
As a result, the higher-order contributions, in the gluonic sector, are
essentially non-abelian.
In particular, the second-loop term should not
contain terms proportional to $C_F^2$, the square of the one-loop
color weight $C_F$. All such terms, present on diagram-by-diagram
level, should eventually cancel, and only the terms proportional
to the essentially non-abelian color weight $C_FC_A$ might remain.
This observation can be used to check the results %
of the higher-order calculations for the relevant %
anomalous dimensions. %

The two-loop contribution to $\Gamma_{cusp}$ in
pure gluodynamics (QCD without massless quarks) was calculated
in our paper \ci{cusp}.%
\footnote{Independently, a similar calculation
          was performed by Knauss and Scharnhorst \ci{Kna}.
          They failed, however, to get rid of all double integrations
          in their result.}
The massless quark contribution can
be also added \ci{IR1} and the total result is
\newcommand\V{\vartheta}
\newcommand\G{\psi}
\ba
\Gamma_{cusp}^{2-loop}(\V,\alpha_\s)&=&
\left(\frac{\alpha_\s}{\pi}\right)^2C_F
\left[-N_f\,\frac5{18}(\V\coth\V-1)
+C_A\left(\frac12+\left(\frac{67}{36}-\frac{\pi^2}{24}\right)(\V\coth\V-1)
\right.\right.  \nonumber
\\
&-&\coth\V\int_0^\V d\G\G\coth\G
+\coth^2\V\int_0^\V d\G\G(\V-\G)\coth\G
\lab{cusp} \\
&-&\frac12\sinh2\V\int_0^\V d\G\frac{\G\coth\G-1}{\sinh^2\V-\sinh^2\G}
 \left.\left.\log\frac{\sinh\V}{\sinh\G}\right)
     \right] .
\nonumber
\ea
It explicitly satisfies the requirement of the non-abelian
exponentiation theorem:
there are no $C_F^2$-terms in it.

In the framework of the HQET,
the two-loop calculation of the velocity-dependent anomalous dimension
was performed in a recent paper by Ji \ci{Ji}.
The author presents his result diagram by diagram,
in terms of 7 integrals $I_1,\ldots, I_7$,
two of which ($I_1$ and $I_7$) are calculated explicitly
and 5 others are claimed to be
not calculable in terms of the elementary functions.
The integral $I_6$ even contains a double integration.%
\footnote{We failed to check this integral,
          since the explicit expression presented in \ci{Ji}
          contains a variable $u$ (or a function $u$) not defined
          in the paper.}
{}From our formula \re{cusp}, it is clear
that there should be really only three
independent integrals containing no double integrations.
The most worrying is the fact that
the $C_F^2$-term in \ci{Ji}
is not explicitly equal to zero: it is
proportional to ($I_1I_3+I_4-I_5$).
Checking the results of \ci{Ji}, we managed to show that
$I_3=0$. This observation allowed us to simplify
another integral, $I_4$. Then, performing a
straightforward integration by parts, we obtained that $I_5 = I_4$.
Thus, the total $C_F^2$ coefficient in \ci{Ji}
vanishes, as it should.


The same statement is valid for the end-point
anomalous dimension $\gamma_{end}(\alpha_\s)$ or, equivalently, for the
anomalous dimension of the effective heavy field in the HQET.
In particular, no $C_F^2$-terms really appear at the two-loop level.
The authors of \ci{Gro}, apparently,  have not realized
that their two-loop results have this property.

\subsection{Cusp anomalous dimension for small and large angles}
Incorporating the exponentiation theorem, one can easily
obtain an important result mentioned in the introduction
that the anomalous dimension $\Gamma_{cusp}(\V, \alpha_\s)$,
for large angles $\vartheta$,
is linear in $\vartheta$ (or $\log(Q^2/m^2)$)
to all orders of perturbation theory:
\be
\Gamma_{cusp}(\V, \alpha_\s) |_{\V \to \infty} =
\V K(\alpha_\s) +  {\cal O}(\V^0) \, .
\lab{large}
\ee
This fact has a simple explanation. In this limit,
$\log(Q^2/m^2)$ is a typical collinear logarithm resulting from
integration over the regions where the gluons
are almost collinear to $v_1$ or $v_2$,
{\em i.e.,} when they are emitted/absorbed at small angle.
However, in the two-particle irreducible diagrams
(only these really contribute to $\Gamma_{cusp}(\V, \alpha_\s)$)
there is just one angular integration producing
a singularity in the $m \to 0$ limit.
Hence, just a single logarithm emerges,
and we get eqs.\re{gammaIR}, \re{large}.

Using the explicit expression \re{cusp},
one can check that our $\Gamma_{cusp}(\V, \alpha_\s)$
is linear in $\vartheta$
for large $\vartheta$. However,
this linearity requirement is not satisfied by the result \ci{Ji}
given by  Ji, since it contains an exponentially rising
contribution ${\cal O}(\exp(\vartheta))$ in the term proportional
to $C_FN_f$. In general, we disagree with his results
for the terms having this color factor.

{}From the expression \re{kalpha}, it follows that
the first two coefficients of the perturbative expansion
for the asymptotic slope $K(\alpha_\s)$
are positive. There are good reasons to expect that
$K(\alpha_\s)$ should be  positive to all orders.
In particular,  it can be shown \ci{IR1} that
$K(\alpha_\s)$ determines the asymptotic
behavior of the parton distribution
function $f(x,Q^2)$
as $x\to 1$ for large $Q^2$.  More specifically, the exponent
$a(Q^2)$ in the asymptotic representation
$f(x,Q^2)\propto (1-x)^{a(Q^2)}$ obeys the
evolution equation \ci{IR1}
$$
\frac{d}{d Q^2}a(Q^2)=K(\alpha_\s(Q^2)).
$$
If $K(\alpha_\s)$ is a positive function, then $a(Q^2)$ is an
increasing function of $Q^2$, and, for $x \sim 1$, the parton
distributions tend to zero
faster and faster with increasing $Q^2.$ And that is
just what one is expecting  within the QCD parton picture:
if one probes the structure of a
hadron at smaller and smaller distances,
the probability to find a parton carrying
almost all momentum of the hadron
should become  smaller and smaller.

In the opposite limit of small cusp angles, eq.(\ref{cusp})
gives
\be
\Gamma_{cusp}(\V,\alpha_\s)
|_{\V \to 0} =
\V^2\left[ \frac{\alpha_\s}{3\pi}C_F+\left(\frac{\alpha_\s}{\pi}\right)^2
C_F\left(C_A\left(\frac{47}{54}-\frac{\pi^2}{18}\right)-
N_f\frac5{54}\right)\right]
+ {\cal O}(\V^4)
\lab{small}
\ee
Apparently, the expansion of
the two-loop cusp anomalous dimension \re{cusp} contains
only even
powers of the  angle $\vartheta$.
In view of the equivalence relation
\re{eqHQET} this seems natural:
$\Gamma_{cusp}(\vartheta, \alpha_\s)$  depends on $\vartheta$ through
$w=\cosh\vartheta$,  which is
an even function of $\vartheta.$
However, this simple reasoning  is true only if
$\Gamma_{cusp}(\V, \alpha_\s)$ is analytic at $\V=0$ or,
equivalently, if
$\gamma_{w}(\alpha_\s)$  is
an analytic function of $w$ in the
vicinity of the point $w=1$. This fact is not immediately
obvious,  since the explicit 1-loop expression (\ref{gammaw1loop})
for $\gamma_w(\alpha_\s)$ contains the square root $\sqrt{w^2-1}$
capable of generating a cut starting just at $w=1$.

To study the analyticity properties of
$\gamma_{w}(\alpha_\s)$, it is useful to use its relation
to the anomalous dimension of the IR factor
$F_{IR}$ (\ref{eqircusp}) and the fact that $F_{IR}$
enters into the factorized expression for
the elastic quark form factor $F(Q^2,m^2).$
Hence, $\gamma_{w}(\alpha_\s)$
possesses only those singularities
which the quark form factor has as a function
of $Q^2=2m^2(w - 1).$
It is well known, that
the nearest singularity of the form factor
is due to the pair creation
in the annihilation channel and it corresponds to the point
$Q^2 =  -4m^2.$
This implies that $\gamma_w(\alpha_\s)$ is
an analytic function at $w=1$,  with the nearest
singularity at $w=-1$.
Indeed, using the explicit expressions \re{brems} and \re{cusp},
one can easily check that the
bremsstrahlung function has a cut starting at $Q^2=-4m^2$
while the two-loop cusp
anomalous dimension has  the only pole
singularity  at  $\cosh\V=-1.$\footnote{We
            are grateful to L.N.Lipatov who has drawn our
            attention to this point}

\subsection{Conclusions}
Thus, we demonstrated
that three important functions, that appear in different
branches of the quantum theory of gauge fields, coincide.
The same function, the cusp anomalous dimension
of a Wilson line,
appears when one studies the IR singularities of
the on-shell form factors or the
velocity-dependent anomalous dimension
in the effective heavy quark field theory.
This is because in both cases, all the IR induced terms can be
factorized from the original amplitudes into the
universal IR factor given by the vacuum average of a Wilson line.
This object, though it is a {\em nonlocal} functional
of the gauge fields, has rather simple (and well studied)
renormalization properties. As a consequence, the
anomalous dimensions of the Wilson lines govern
the IR logarithms in the
same manner as the anomalous dimensions of
the {\em local} composite
twist-2 operators control the collinear
singularities one encounters studying the structure
functions of deep inelastic scattering.

In practical aspect,  the well-developed machinery of the Wilson
loop formalism can provide a good support for the
heavy quark effective field theory.
In particular,  we established the equivalence of some recent
HQET calculations \ci{Uni,Gro} with the earlier results
for the Wilson lines \ci{Pol,peo,Kna}. We also corrected the
two-loop calculation \ci{Ji} of the velocity-dependent anomalous
dimension.

\bigskip
{\em Acknowledgment. }
One of us (A.V.R.) is grateful to E. De Rafael and
J. Soffer for the warm  hospitality at CPT (Marseille).
G.P.K. thanks G.Marchesini for useful discussions and
support.

\bb{99}
\bi{Vol} %
        M.B.Voloshin and M.A.Shifman, Sov. J. Nucl. Phys.
        45 (1987) 292 ; 47 (1988) 199.
\bi{Isg} %
        N.Isgur and M.Wise, Phys. Lett.  B232 (1989) 113;
        B237 (1990) 527.
\bi{Eich} 
        E.Eichten and B. Hill, Phys. Lett. B234 (1990) 511.
\bi{Georgi}
        H.Georgi,  Phys. Lett. B240 (1990) 447.
\bi{Uni} 
        A.F.Falk, H.Georgi, B.Grinstein and M.B.Wise,
        Nucl. Phys. B343 (1990) 1.
\bi{Pol} 
        A.M.Polyakov, Nucl. Phys. B164 (1980) 171.
\bi{peo} 
        V.S.Dotsenko and S.N.Vergeles, Nucl. Phys. B169 (1980) 527;
\\      R.A.Brandt, F.Neri and M.-A.Sato, Phys. Rev. D24 (1981) 879;
\\      H.Dorn, Fortschr. Phys. 34 (1986) 11.
\bi{IR}  
        S.V.Ivanov, G.P.Korchemsky and A.V.Radyushkin,
        Sov. J. Nucl. Phys. 44 (1986) 145;
\\      G.P.Korchemsky and A.V.Radyushkin, Phys. Lett. 171B (1986) 459;
        Sov. J. Nucl. Phys. 45 (1987) 127; \, 910;
\\      G.P.Korchemsky, Phys. Lett. 217B (1989) 330; 220B (1989) 629.
\bi{IR1} 
        G.P.Korchemsky, Mod. Phys. Lett. A4 (1989) 1257.
\bi{Col} 
        J.C.Collins, in ``Perturbative Quantum Chromodynamics'',
        ed. by A.H.Mueller (World Scientific, Singapore, 1989) p.573.
\bi{D-Y} 
        C.T.H.Davies and W.J.Stirling, Nucl. Phys. B244 (1984) 337;
\\      W.L. van Neerven, Phys. Lett. 147B (1984) 175.
\bi{x->1}
        J.Kodaira and L.Trentadue, Phys. Lett. 112B (1982) 66.
\bi{Fra} 
        D.Yennie, S.Frautschi and H.Suura, Ann. Phys. (N.Y.)
        13 (1961) 379; \\
        G.Grammer and D.Yennie, Phys. Rev. D8 (1973) 4332;
\\      J.M.Cornwall and G.Tiktopoulos, Phys. Rev. D13 (1976) 3370;
        D17 (1980) 2837.
\bi{Ger} 
        J.Gervais and A.Neveu, Nucl. Phys. B163 (1980) 189;
\\      I.Ya.Aref'eva, Phys. Lett. 93B (1980) 347.
\bi{Kna} 
        D.Knauss and K.Scharnhorst, Ann. der Phys. 41 (1984) 331.
\bi{Gro} 
        D.J.Broadhurst and A.G.Grozin, Phys. Lett. 267B (1991) 105.
\bi{Gat} 
        J.G.M.Gatheral, Phys. Lett. 113B (1984) 90;
\\      J.Frenkel and J.C.Taylor, Nucl. Phys. B246 (1984) 231.
\bi{cusp} 
        G.P.Korchemsky and A.V.Radyushkin, Nucl. Phys. B283 (1987) 342.
\bi{Ji} 
        X.Ji, Phys. Lett. 264B (1991) 193.
\eb

\end{document}